# Analysis and Monte Carlo Modelling of Multiple Scattering of a Beam of Neutrons or Photons off of the Two Surfaces of a Broad Material Target as a Function of its Thickness

Eric V. Steinfelds
Keith Andrew
Department of Physics and Astronomy
Applied Physics Institute
Western Kentucky University, Bowling Green, KY 42101

Abstract

We consider the scattering of neutrons and photons on solid volume rectangular targets. It is common to treat this problem using the Maxwell Boltzmann Transport Equation and to use underlying symmetries to simplify the calculation. For isotropic scattering centers we can introduce a direct Fredholm integral equation approach to finding the flux. Here we compare a Monte Carlo evaluation of the resulting Fredholm equation to a deterministic iteration method of solution. We include a kernel based method for estimating the overall angular dependence. We find that for simple geometries utilized in studies of radiative dosimetry, of neutron shielding assessments, and indirectly of criticality that we get reasonably rapid convergence of flux and current values.

I. INTRODUCTION

There are many interesting problems to solve in radiation transport theory relevant to nuclear engineering, to the design of radiation therapy devices, and to analysis and modeling of planetary and astrophysical systems. A particularly relevant endeavor to these various fields, is the systematic and ideally explicit prediction of the spatial distribution, via various processes, of radiation flux and the angular distribution of radiation flux. Three-dimensional transport methods have been used to track and monitor types of dosimetry[1] and generalized for a variety of rough and semi-smooth surfaces[2]. Several methods have been developed with discrete coordinates[3] or for anisotropic scattering[4] using a path linking methodology[5] which can be coupled to gamma ray emission[6]. Neutron transport for the basic slab[7] geometry[8], the general disk geometry[9], finite bounded systems[10], and infinite homogeneous systems[11] have been studied in detail. Photon and neutron transport also arises in an astrophysical[12] setting for star[13], radiation[14] and jet[15] transport where Monte Carlo (MC)[16,17] methods for transport Green's functions[18,19] can be used to evaluate the spatial distribution of photons[20]. In a fashion analogous to neutron scattering the back-scatter of soft photons in problems involving the isotropic scattering of light or with a $\cos(\theta)$ weighting[21] has also been treated analytically while fundamental reactor[22] transport has been modeled using neutron transport[23] Green's functions. In this paper, we provide a description of an analytical serial benchmark for the formula for the scattering distribution of neutrons or of ad hoc isotropically scattered γ-rays from slab shaped and fixed volume targets of thickness ('Tk') subjected to a direct beam of neutrons or 'hard' photons.

Here we will refer to the scalar particle flux as, $\Phi(x, y, z)$, and the angular flux, $\varphi(x, y, z, \theta, \psi)$, by the single character $\varphi$. Indeed, finding the scalar flux from neutron diffusion[24] Green's[25] functions rather than the angular flux from the Maxwell Boltzmann Transport Equation (MBTE) has long been an approved practice in analysis[26] of reactor cores[27], criticality[28], shielding[29], spent fuel[30], and for a variety of difficult to treat amorphous samples[31]. However, there are examples, especially in accelerator systems and on collimators of hard X-rays, where a monodirectional beam of radiative particles attempts to penetrate a target with the effect of the monodirectional distribution of $\varphi$ evolving into a multi-directional distribution of $\varphi$, such that in thick enough targets $\varphi$ can almost be approximated as $\Phi/4\pi$. However, close to the boundaries of entry and escape of the radiative particles (e.g. of neutrons or photons), $\Phi$ by itself gives part but not all of the story of the current density for radiative particles multiplied by the scattered escaping particles. As understood from Fick's 'Law' the diffusion equation does not fully suffice for the beam condition of initial mono-directionality and precise evaluation of the material to vacuum (or air) interface. The completely formal and foolproof way to find the flux and current of particles which scatter from the original beam off the slab or similar fixed volume target is to solve the MBTE, such that the main initial/boundary condition is the approach given by an external primary beam of particles into the target from a remote source in vacuum. If we presume that all the scattering centers encountered by the beam within the target material are isotropic, then the MBTE can be transformed into the following Fredholm integral equation[32][33]:

$$\Phi_{SC}(\vec{x}) = \frac{1}{4\pi} \int_{vol} \left[ \Phi_{SC}(\vec{x}') + \Phi_{inc}(\vec{x}') \right] \frac{\Sigma_s \exp(-\Sigma_t |\vec{x} - \vec{x}'|)}{(|\vec{x} - \vec{x}'|)^2} d\vec{x}' \qquad (1)$$

where $\Phi_{sc}$ is the scattered angular flux, $\Phi_{in}$ is the incident external particle source (i.e. of the beam of irradiation), $\vec{x}'$ represents the spatial extent of the intersection region, and $\Sigma_s$ and $\Sigma_t$ are macroscopically the total cross section and the scattering cross section, respectively. Note that $\Sigma_U$ is the product of the number density of the material times $\sigma_U$ resulting in units of inverse length. This Eq. (1) can be solved with much more ease than the MBTE can be solved, especially if one or two spatial symmetries can be imposed. Moreover, there are known analytical and semi-analytical benchmarks for transport theory with solved cases of Eq. (1) and more elaborate variants for numerous choices of boundary conditions[34]. The Milne-Schwarzschild Integral Equation, which has been studied by planetary physicists, overlaps Eq. (1) except that one of the spatial dimensions usually or always is infinite in the Milne-Schwarzschild Equation[35]. By way of the judicious use of cylindrical coordinates, it is acceptable to approximate a broad square shaped slab with a non-tapered disk for an attenuator (and/or scattering material). In doing so, the integral on the R.H.S. of Eq. (1) undergoes the following simplification for its infinitesimal volume (d$V$)→ ($2\pi$ r) dr dz . By taking advantage of this representation of the 'infinitesimal' volume, Eq. (1) can be simplified from a relation with a triple integral to a Fredholm integral equation whose integral expression involves a single integral and incorporates an exponential integral, as we see here:

$$\Phi_{SC}(x, y, z) = \frac{1}{2} \int_0^{Tk} [\Phi_{SC}(\vec{x}') + \Phi_{inc}(\vec{x}')] \Sigma_s \, E_{[1]}(\Sigma_t |z - z'|) \, dz' \qquad (2)$$

$$\text{where} \quad E_n(x) = \int_1^\infty \frac{e^{-xt}}{t^n} \, dt \quad \text{so} \quad E_1(x) = -Ei(-x) \quad .$$

There are several computational approaches which the applied nuclear physicist or research nuclear engineer can take to properly simulate what comes from the MBTE:

(1) to write one's own MC code which is tailored to the geometry specific to the problem to solve,

(2) to use an established MC code for transport and accordingly selecting or modifying the cross-section library of that code,

(3) to find an analytical solution to the MBTE with numerical assistance in the last few steps; and,

(4) to use a deterministic code in which one attempts to fit the boundary conditions to the input or mesh generator of the deterministic code available to the user.

Codes such as DOORS3[36], TORT[37], and ATTILA[38] come to mind for approach (4). The method of analysis of radiation escape from containers as developed by Steinfelds and Prelas[39] has offered some inspiration for approach (3).

## II. PURPOSE, FEATURES, & CONCEPTS

Here we have produced a MC code to conform to our preferred isotropic scattering cross-sections. We also have developed two thoroughly mathematically based formulations manifested in two complimentary calculation based numerical codes to predict (a) the redistribution of flux due to scattering; (b) the remaining intensity of the primary beam; and (c) the local intensities and angular dependence of the scattered radiative particles. Thus, we have chosen the first (1) and third (3) computational approaches at this time for our endeavor to effectively predict beam scattering phenomena. For our approach to (1), our small-scale MC code is referred to as the Stochastic Multiple Scattering Kernel Engine, or SMuSKE. Our dual tandem deterministic code is referred to as "Secondaries Iteration Enveloper of Flux" (SIEF) and the "Kernel nModulated Angle Map Agent" (KnAMA), for our approach to (3) and can be expressed as:

$$KnaMa\_formula: \quad \langle \cos \theta \rangle = \frac{Hcosth(\Sigma\ )}{H(\Sigma\ )} \qquad (3a)$$

$$H\cos th(\Sigma_{transmit}) = \int_0^{Tk} \int_0^\infty \frac{\Sigma_{SC} \, 2\pi \, r \cos \theta' \, \Phi_{iterated}(z') \, e^{-\Sigma_t \sqrt{r^2 + z'^2}}}{(r^2 + z'^2)^{3/2}} \, drdz' \qquad (3b)$$

$$H\cos th(\Sigma_{transmit}) = \int_0^{Tk}\int_0^{\infty} \frac{\Sigma_{SC}\, 2\pi r\, z'\, \Phi_{iterated}(z')\, e^{-\Sigma_t \sqrt{r^2+z'^2}}}{(r^2+z'^2)^2}\, drdz' \qquad (3c)$$

$$H(\Sigma_{transmit}) = \int_0^{Tk}\int_0^{\infty} \frac{\Sigma_{SC}\, 2\pi r\, \Phi_{iterated}(z)\, e^{-\Sigma_t \sqrt{r^2+z^2}}}{(r^2+z^2)^{3/2}}\, drdz \qquad (3d)$$

Eq. (3b), for the case of neutrons ($z_{[observe]}=Tk$) on the transmission side, has $\cos(\theta)$ equal to z' divided by $(r^2+z'^2)^{1/2}$. Our deterministic dual, SIEF and KnAMA calculator-codes, can be run independently or set in a cross-iteration mode with each other. KnAMA has a low (but not zero) sensitivity to SIEF. The first primary goal of this work is to demonstrate the accuracy and effectiveness of the Kernel nModulated Angle Map Agent (KnAMA) as a predictor of the direction of the escaping current from the main forward and backward surfaces of the fixed volume target for an incoming radiation beam. The second goal is to predict the transmission intensity of the particles. Additional goals are to predict the spatial distribution of dose inside of the target by simply using

$$\Phi_{scat[N+1]}(0,0,z) = \frac{1}{2}\Sigma_s \int_{vol} \Phi_{scat[N]}(0,0,z')\, E_{[1]}(\Sigma_t |z-z'|)\, dz' \qquad (4a)$$

$$\text{where}\quad \Phi_{scat[1]}(0,0,z) = \frac{1}{2}\Sigma_s \int_{vol} \Phi_{beam}(z',\Sigma_{totl})\, E_{[1]}(\Sigma_t |z-z'|)\, dz' \qquad (4b)$$

$$\text{and}\quad \Phi_{inc} = \Phi_{beam}(z,\Sigma_{totl}) = \Phi_{beam}(z=0,\Sigma_{totl})\exp(-\Sigma_{totl} z)\ ,\ straight, \qquad (4c)$$

$$\text{where}\quad \Sigma_{totl} = \Sigma_{sc} + \Sigma_{abs}\ .$$

Our iterative method of solving for the scattered flux in the SIEF program enables us to find $\Phi(x, y, z_{inside})$ and $\Phi(x, y, z_{surface})$ inside of and directly on the vacuum boundaries of the target. The predictions of the position dependent flux generated by the "Secondaries Iteration Enveloper of Flux" (SIEF) and of the average angle of escape of particles from the KnAMA calculator enable us to predict the current density, $\mathbf{J}(x)$, of those particles which penetrate (and thus forward escape) a broad rectangular barrier or cylindrical barrier and of those particles which get back-reflected from the rectangular barrier. These outwardly wandering particulate current densities are given by:

$$J_{trans} = \frac{Hcosth(\Sigma)\Phi_{iterated}(z=Tk)}{H(\Sigma)}\Big|_{z=0}^{z=Tk} \qquad (5a)$$

$$|J_{refl}| = \frac{Hcosth(\Sigma)\Phi_{iterated}(0,_{otherWay})}{H(\Sigma)}\Big|_{z=Tk}^{z=0} \qquad (5b)$$

.

The profile of trajectories of individually back-reflected particles (i.e. of (5b)) looks somewhat similar to the profile of trajectories of visible photons which undergo diffuse reflection from a slab of shiny but unsmooth marble.

In this paper and in the initial formulations within SIEF and KnAMA, we have presumed that the internal atoms and nuclei are purely isotropic scatterers of neutrons and hard photons (or γ-rays or X-rays). Our SMuSKE particle modeler is designed to stochastically execute isotropic scattering, rather then anisotropic scattering but SMusSKE has options for adjusting and setting the parameters for albedo ratio and others. The adjustable albedo ratio, R, is the ratio of the respective microscopic cross-sections of scattered (sc) and of total (t) particles:

$$Albedo = 1 - \frac{\sigma_{abs}}{\sigma_t} = R$$

$$R = \frac{\sigma_{sc}}{\sigma_t} \quad where \quad \sigma_t = \sigma_{abs} + \sigma_{sc} \quad . \tag{6}$$

The SMUSKE program user can adjust 'R', the number density of atoms in the target slab, the angle of entry of the particles in the mono-directional radiative beam, and the number of neutrons or photons in this source beam. In all the simulations done for immediate examples, we selected a beam target of finite thickness along the z-axis and infinite breadth along the x and y axes.

## III. SIMULATIONS, EXAMPLES, & RESULTS

We have established the analytical calculations of the first iteration of $\Phi_{scat[N]}$ for our calculations of $\Phi(z)$ via the deterministic method of SIEF. If we follow from our Eqs. (4), we get:

$$\Phi_{beam}(z,\Sigma) = \Phi_{beam}(0,\Sigma)\exp(-\Sigma_{tot} z), \text{ of normal beam.}$$

$$\Phi_{scat[1]}(z,\Sigma) = \Phi_{beam}(0,\Sigma)\frac{1}{2}\frac{\Sigma_{sc}}{\Sigma_{tot}}\left[E_{[1]}(\Sigma_{tot}\, z) - E_{[1]}(\Sigma_{tot}(z-Tk))\exp(-mfpm) + g(z,\Sigma_{tot})\exp(-\Sigma_{tot} z)\right] \tag{7}$$

$$where \quad g(z,\Sigma_{tot}) = \left(\gamma + \ln 2 + \ln(\Sigma_{tot} z + E_{[1]}(\Sigma_{tot} 2(Tk-z)))\right)$$

where γ gamma is the Euler Mascheroni constant. In Eq(7), we often rewrite a factored expression (T $\Sigma_{tot}$) as 'mfpm', or mean free path multiple. As the experienced applied mathematician might intuit and quickly discern the explicit formula for $\Phi_{scat}[2](z,\Sigma)$ would barely fit within one page of text. Likewise, the formula for $\Phi_{scat}[3](z,\Sigma)$ would need more than 2 pages of text. With computational methods, we did compute $\Phi_{scat[2]}(z,\Sigma)$ through $\Phi_{scat[8]}(z,\Sigma)$ for various choices of mfpm by way of this method of iteration and for use in the SIEF 'calculator' engine. Note: beware of the following mnemonic inequivalence: 'mfpm' ≠ m.f.p.

In Table I, the predictions of KnAMA are given for the average direction of all scattered radiative particles present within the current density which escapes through the reflective (i.e. rear) surface and which escapes through the forward (i.e. remotest along z-axis) surface. Likewise, the respective results of average particulate direction generated by SMuSKE are given. The quantities in columns 2 through 5 are the average values of cos($\theta_{escape}$) per escaping particle. A study of transport theory and a review of the Gauss Law application for the current sources reveals that the if the slab has isotropic scatterers, then the statistically expected value of cos($\theta_{escape}$) at a flat interface into vacuum needs to be cos($\theta_{<escape>}$) ≥ ½, where the expected angular value of escape matches $\theta_{escape}$ ≤ 60º. Note: $\theta_{escape}$ =0º means that the neutron travels in a normal ray from the surface of the interface. These values for θ fit exactly with the pattern that we specifically see in rows 2, 3 4, and 5 of Table I. In Tables I and II, we have set the albedo ratio equal to 1.00, (i.e. zero absorption). The symbol 'mfpm' means total mean free path multiple of $\Sigma t$ when mfpm is mentioned in relation to Eq. (7). The values of mfpm are in the 1st column of Tables I and II. 'Tk' is the thickness of the target. For example, if 'Tk' of the slab equals 4× total Mean Free Path, then 'mfpm' equals ¼.

TABLE I

| mfpm | Ave. cos($\theta_{esc}$) backward | | Ave. cos($\theta_{esc}$) forward | |
|---|---|---|---|---|
| | KnAMA backward | MC* backward | KnAMA forward | MC forward |
| $10^{-7}$ | 0.5000 | --- | 0.5000 | --- |
| 1/40 | 0.517891 | Low rate | 0.518012 | Low rate |
| 1/12 | 0.539633 | 0.56468 | 0.540493 | 0.57507 |
| 0.30 | 0.58198 | 0.55756 | 0.583445 | 0.57656 |
| 0.60 | 0.61002 | 0.58488 | 0.61704 | 0.60217 |
| 3.0 | 0.66007 | 0.63116 | 0.701 | .680598 |
| 5.0 | 0.63738 | 0.62546 | 0.757306 | .68526 |
| 6.5 | .63759 | 0.61769 | 0.773951 | .68985 |

*For low mfpm values, the population is set ≈ 2000.

In Table II, ratios of the (escaping scattered current plus non-attenuated current densities) over the incoming current densities are given in columns 2 through 5. The Monte Carlo tool chosen here is SMuSKE. In Table II, "MC forward" refers to SMUSKE.

TABLE II

| | Ratios of all out-going $|\vec{J}|$ over $|\vec{J}_{[incoming]}|$ @ | | | |
|---|---|---|---|---|
| mfpm | SIEF backward | MC backward | SIEF forward | MC forward |
| $10^{-7}$ | 0 | --- | 1 | 0.999 |
| 1/40 | 0.004 | lack data | 0.996 | lack data |
| 0.05 | 0.005550 | 0.01502 | 0.993713 | 0.9853 |
| 0.3 | 0.064441 | 0.1101 | 0.928732 | 0.8905 |
| 0.6 | 0.164415 | 0.20667 | 0.835254 | 0.79333 |
| 3.0 | 0.633700 | 0.65134 | .3612993 | 0.34866 |
| 5.0 | 0.81843 | 0.81428 | 0.181241 | 0.18571 |
| 6.0 | 0.901983 | 0.89285 | 0.118016 | 0.10714 |
| 6.5 | 0.84674 | 0.90957 | 0.153254 | 0.09042 |

@ note: $\vec{J}_{[incoming]}$ is current density of inc. beam.

If the absorption ratio (that is 1─albedo ) were equal to zero, it would be interesting to add a sixth column to Table II, in which the attenuation coefficients of extreme absorption are included, where the scattering $\sigma_s$ = 0. Corresponding to 'mfpm' values of {0.05, 0.3, 0.6, 3.0, 5.0, 6.0, 6.5} we classically obtain the following factors of remote escape: {0.951 ,0.7408, 0.549, .0498, .0068, .0015}. The significance of this exponentially declining sequence of numbers is that it shows how much larger the transmission through the target slab is when non-absorptive scattering of the neutrons or photons occurs in the target material than if pure absorption per collision were to occur in Table II.

## IV. CONCLUSION

The concise analytical predictions of the relevant macroscopic current densities are the KnAMA based predictions of |**J**(x,y,z=0)| and **J**(x,y,Tk), which are inspired by the Green's function which is present within the kernel on our KnAMA calculator. These concise analytical predictions of the relevant macroscopic current densities demonstrate accuracy and effectiveness due to the ability of the KnAMA calculating code and the SMuSKE simulator to predict the average cos(θ) values of escape to reasonably close agreement. The CPU time of a KnAMA calculation is more than 20 times faster than a SMuSKE (truly stochastic) simulation per example. In Table I, we see that the prediction of the average cosine of angle of escape of a particle backwards agrees within 7% to that of the MC Simulation. When the thickness (Tk) of the target is less than 3 times greater than (i.e. when mfpm ≤ 3) the total mean free path of radiative particles (n`s or γ`s), we find the agreement between the two independently generated predictions converges to within 5%. The reason for fairly close agreement rather than extremely close agreement is due both to the statistical difficulty of acquiring enough scattered particles for angle tallies when M.F.P[tot], or Mean free Path[tot], is ≥ 8 Thickness (i.e. 8 Tk). This unreliability will be addressed and noticeably diminished by extending a polynomial and increasing the number of iterations for the Math Kernel Function within SIEF. In consideration of this issue of the numerical predictions of SIEF, we inspect Table II, which mostly uses SIEF for the deterministic calculations in columns two and four. KnAMA is only a minor assistant to SIEF in (II). We see in the very bottom row of Table II a rather mediocre agreement between the Monte Carlo and SIEF predictions of escape ratios. This is because the iterative math kernel function within SIEF is at least two terms short of what is quantitatively needed when 'mfpm' is set equal to 6.5. This issue has been corrected by the adding of 3 extra iterations upon the kernel function for the computation of the case when mfpm=6 for the 2nd to last row of (II). Accordingly, the MC and revised SIEF predictions agree quite closely in the case when mfpm=6. In March 2014, the upgrade in iterations was invoked in SIEF for all cases comparable to those of Table II where mfpm ≥ 3 and albedo ratio is ≥ 0.6. Due to (A) the fact that SIEF has not given any absurd predictions such as over estimating the amount of multiple scattering, (B) to the fact that the first iteration for Φ in SIEF (given in Eq. (7)) is on solid mathematical footing, (C) the fact the first iteration of scattered Φ shown in Eq. (7) demonstrates a correct pattern contribution to flux of a perturbative nature, and (D) due to the fact that the iterations of $\Phi_{scat_{[N]}}(0,0,z)$ from Eq.(4) asymptotically evolves as approximately a geometric series with respect to 'N' lead the authors to conclude as a mathematical conjecture that SIEF is potentially very reliable for future constructions of improved versions of Table (II) . We assert that iterative formulations and enumerations of Φscat[11], Φscat[12], through Φscat[20] will very likely further vindicate, but not per se formally prove, the conjecture that that SIEF is incrementally reliable for thick slabs of scattering material as the number of Φscat[N] terms increase. [40]

There is a benefit to having analytical benchmarks for the sake of comparisons for results from MC simulations and computationally efficient (but generic) established deterministic codes. KnAMA can now be demonstrated to have 5% agreement with our MC based SMuSKE code, even when the distribution of Flux(x,y,z) is skewed at first guesses in the input of KnAMA. KnAMA can calculate the average angle of current in less than one twentieth of the time that a typical MC code such as KnAMA takes when the standard deviations of the simulation are comparable to the numerical uncertainties in the formula expansions used. KnAMA has an analytical robustness, completeness (for subcritical fuels and later crit. fuels), which gives it a major advantage for predicting average angles. KnAMA is rather specialized and lacks the versatility in geometric shapes and accessible choices of materials which accompany some of the MC transport codes[41].

However, when it comes to predicting the average angle of escaping current of rectangular slabs and single macroscopic spheres, KnAMA is much faster than established/semi-established MC codes and gradually will come to equal the accuracy of these established (or semi-established) MC transport codes such as EGSnrc and GEANT4 [42] for calculating the angular distribution of current from the macroscopic targets with purely isotropic atoms/nuclei and the geometry(ies) discussed in this paper. Of course, the established MC codes such as EGSnrc and GEANT4 are not limited to isotropically distributed nuclei and atoms for targets. Nevertheless, for materials such as lead for a slab, it is acceptable for some neutrons and keV range photons to presume approximate isotropic scattering. Accordingly, our SIEF 'calculation kernel' and our SMUSKE modeler both are formulated with target atoms and/or nuclei which are isotropically distributed scattering centers. In the future the authors intend to enhance KnAMA with an attached data base of internal scalar flux map approximations of $\Phi(x,y,z)$ in order to take advantage of the relatively small sensitivity of $\cos(\theta)$ to the exact shape of $\Phi(x,y,z)$. KnAMA generates the prediction of angle extremely quickly, and the mathematical Kernel of KnAMA can be written out within one page with good detail - if we use a simply fitting yet sufficiently robust approximation for $\Phi$ as input for Kernel of KnAMA.